 \documentclass[11pt,floatfix]{article}

\usepackage{graphicx}% Include figure files
\usepackage{setspace}
\begin{document}

\begin{center}%\setcounter{page}{0}..\newpage
\huge{\bf The biological frontier of pattern formation} \\[5mm]
%\normalsize
\normalsize L.~M. Pismen\\
\emph{Technion -- Israel Institute of Technology, Haifa 32000, Israel}
\end{center}
%\maketitle
\vspace{5mm}
%\begin{spacing}{1.5}
\noindent
Morphogenetic patterns are highly sophisticated dissipative structures. Are they governed by the same general mechanisms as chemical and hydrodynamic patterns? Turing's symmetry breaking and Wolpert's signalling provide alternative mechanisms. The current evidence points out that the latter is more relevant but reality is still far more complicated.

\vspace{5mm}

\section{Dissipative structures: from hydrodynamics to chemistry}

Ilya Prigogine coined the term \emph{dissipative structure} \cite{Prigogine,PrigogineN} to bring under a common roof the phenomena that have always fascinated people but only marginally touched the mainstream science. First scientific studies of what would be encompassed by this term, in Faraday's vibrating liquid layers \cite{Faraday} and B\'enard convection \cite{Benard} originated in fluid mechanics. This was an inspiration and example but it was natural and non-controversial; after all, everybody is used to a play of see waves and river currents. Perhaps only the regularity of patterns, the hexagonal tiling resembling the crystalline order that maintains itself far from equilibrium calmness, that was unusual and thought provoking.

It was not hydrodynamic but chemical structures that excited the pioneering work of mid-20th century, since they promised to show the way to understanding the basic mechanism of morphogenesis  \cite{Turing} and evolution \cite{Eigen,evolution}. The paper bearing the ambitious title ``The chemical basis of morphogenesis" \cite{Turing} (winged by the fame of the Turing machine and Enigma Code but more cited than read) ends on a humble note: ``It must be admitted that the biological examples which it has been possible to give in the present paper are very limited. This can be ascribed quite simply to the fact that biological phenomena are usually very complicated".  

The rational message to be extracted from the 36 long pages is that the formation of chemical patterns requires, in the simplest setting, combining a slowly diffusing activator with a rapidly diffusing inhibitor. This principle, that can be established in few lines by linear stability analysis of a two-component reaction-diffusion system, is prominent in the model pattern-forming systems \cite{FitzHugh,Nicolis}, and has  been later clearly formulated in a general form \cite{Segel}. General problems become easy in retrospect; even nonlinear features of model patterns can be explored analytically using the scale separation between the diffusional lengths of the two species \cite{P06}, but particular realisations of general principles overgrow by details like a ship's hull by barnacles.  

Much attention has been paid to the Belousov--Zhabotinsky (BZ) reaction displaying dazzling chemical waves \cite{BZ}; much later, stationary patterns have been also observed in precision experiments\cite{dekepper,Sweeney}. Patterns and waves on a finer scale and with added anisotropy, observed on catalytic surfaces, earned the Nobel prize to Ertl \cite{Ertl}. Simple reaction-diffusion systems served as reasonable, if not quantitative, models of these patterns, even though the precise mechanism of the BZ reaction is disputed to this day. The same models were successfully applied to more complicated phenomena, such as desert vegetation patterns \cite{meron}. To be fair, generic models, such as the complex Ginzburg--Landau \cite{AK} or Swift--Hohenberg \cite{SH} equations reproduce qualitatively the same variety of patterns. Generalised models of this kind are equally applicable to chemical, hydrodynamical, optical, and population patterns \cite{CH93}. But does the Turing mechanism indeed qualify as the chemical basis of morphogenesis?

\section{Wolpert \emph{vs.}\ Turing}

There is a strong general reason why the answer should be negative. The observed chemical patterns, as well as patterns in model computations, are repetitive, and their apparent variety is caused either by a difference in initial conditions or by random inputs. On the contrary, the morphogenetic process is unique and precise, with variegated features emerging at precise locations. 

A rival morphogenetic scenario, supported by the evidence accumulated during half a century \cite{nusslein,Schier,Wolpert16}, has been put forward by Wolpert \cite{Wolpert}. Unlike Turing patterns emerging on a homogeneous background, patterns of biological development are governed by morphogens emanating from a certain source, thereby breaking the symmetry of a featureless background, and the positional information is provided by \emph{morphogenetic gradients}. 

Repetitive pattern do occur; for example, the formation of fingers has been reproduced by a rather realistic two-dimensional simulation \cite{fingers}. Fingers have, however, to be generated at a proper location, which requires positional information, and in a predetermined number, which requires scale invariance. Moreover, all fingers are different, which should depend on positional information along their sequence, Even in development processes generating repetitive regular patterns, such us segmentation \cite{sgmn}, hair follicle or feather formation \cite{sick06,lin09}, or the development of ommatidia in the \emph{Drosophila} eye \cite{Greenwood,Kumar}, these patterns do not emerge on a homogeneous background and triggered by random fluctuations but are generated  by a morphogenetic wave propagating in a predetermined direction. As an extreme demonstration of a contrast between a generic Turing-like model and biological reality, one can compare superficially realistic animal coat patterns modelled by solving the FitzHugh--Nagumo equations \cite{Murray} with the actual intricate three-layer mechanism \cite{colour}.

The Turing and Wolpert scenarios do have common features. Combining activating and repressing agents is necessary in both scenaria, and the essential feature of acting genetic schemes is a \emph{feed-forward} motif  \cite{motif}. The simplest patterning scheme involves a single incoherent feed-forward loop  $S \to P,\; S \to T, \; P \dashv T$ that includes two activating ($\to$) links with different thresholds initiated by the same signal $S$ (induced by a morphogen $M$), and an inhibiting ($\dashv$) link from the intermediate protein $P$ to the target. This scheme generates the classical ``French flag'' pattern \cite{Wolpert} with the target $T$ expressed in the middle (``white'') interval, where the signal level is below a higher threshold of the link to the protein $P$ and above a lower threshold of the direct link to the target. Differences in diffusivities of morphogens also play a role in localising activating or repressing thresholds, although there is no general reason for the latter being less diffusive. And, of course, all morphogenetic patterns are dissipative structures in a wide sense, as they are actively driven and sustained far from equilibrium.   

The ``French flag'' scheme can be straightforwardly extended to two dimensional (2D) patterns under the combined action of crossed gradients. A common example of 2D signalling is found in combined anterior-posterior and dorsal-ventral gradients in a developing \emph{Drosophila} eggshell \cite{Rho_expr2}. A second signal, generated by a morphogen with the gradient in the direction normal to that of the first signal, may induce 2D patterning that can be presented, extending Wolpert's simile, as the nine-colour superposition of French and German flags \cite{P11}. For example, expression in the middle domain can be achieved by combining two incoherent feed-forward loops, provided both excitation links are necessary for expression of the target $T$  (AND logic). 

This is, however, far from sufficient to explain the rich variety of locations and shapes of expression domains. Notwithstanding the complexity of  intracellular interaction schemes, the variety of persistent expression patterns cannot exceed the limit set by intersections of two level sets of the two signals. Modifying the form and location of the signal source, e.g. replacing a linear source by a point one, would change only the shape but not the topology of the expression domains. Adding more initiating links with different thresholds may increase the number of subdivisions but domain shapes will be always set by the signal level sets, making it difficult to explain less regular gene expression patterns, such us cusp-like or ``eyebrow"-shape groups of cells in the \emph{Drosophila} eggshell \cite{Rho_expr1}. 

 \section{Localisation, scaling, and robustness}
 
A possible way to creating variegated shapes of expression domains is combining external signals with \emph{autocrine} morphogenetic signalling initiated within the embryonic tissue by ligands whose expression, in turn, is determined by the local morphogen levels \cite{P11}. If there is a single target gene and a single ligand, there are altogether 16 combinations of target and ligand expression in the presence or absence of the autocrine signal. The diffusional range of the autocrine ligand is typically shorter than that of externally supplied morphogens. If, for example, a particular genetic network is built in such a way that in one of the nine domains (say, A) of the ``Franco-German flag" the target is not expressed but the ligand is expressed, while in a neighbouring domain B the ligand is not expressed and the target is expressed only in the presence of the autocrine signal, the target expression will be observed only in a narrow strip near the boundary where the level of the autocrine signal is sufficiently strong. Conversely, if in the domain A both the target and the ligand are expressed, while in the domain B the ligand is not expressed and the target is expressed only at a low level of the autocrine signal, the target will be expressed everywhere except in a narrow strip in the domain B near the boundary with the domain A. This helps to explain gene expression in domains of convoluted form, e.g. that involved in the formation of dorsal appendages in the \emph{Drosophila} eggshell \cite{dorsal}. The number of  combinations grows exponentially as $2^{2(n+1)}$ with the number $n$ of  autocrine ligands, leading to a great variety of expression domains for the same intrinsic genetic scheme.

The scaling and robustness problem is an Achilles heel of both Turing patterns and Wolpert's flags, and it is not helped at all by adding more signalling species. In both cases, the scale of a pattern is fixed by diffusional lengths of, respectively, reactants or morphogens, while in reality the development patterns are scaled by the size of an organism; as an extreme example, a mouse and a giraffe have the same number of vertebra. A variety of mechanisms were suggested to rectify this contradiction. Some naive attempts suggested doubling a signal by either a counter-propagating signal or a sink at the opposite edge -- an arrangement becoming forbiddingly clumsy in the case of 2D patterning and in the presence of several morphogens and never supported by evidence. A more rational recipe \cite{Barkai2} is to enhance the decay rate of a morphogen at higher concentrations by making it dependent on the concentration level. This adds robustness by buffering fluctuations at the source. The scaling problem can be solved by some kind of global control \cite{Barkai}, which is also known to stabilise localised structures in model reaction-diffusion systems \cite{KM,P94}. For example, making the morphogen degradation dependent on some chemical species present in a fixed amount and uniformly distributed in a developing embryo would automatically make the morphogen gradients scale-invariant; it could also act in the same way on all morphogens diffusing in different directions. Global agents require, however, a fast mechanism for sustaining their uniform concentration which cannot be realised by molecular transport. Long-range mechanical interactions may, in principle, serve as a mechanism of global feedback \cite{Belintsev} but the actual mechanism is likely far more complicated. 

The efforts to understand morphogenetic problems in the diffusional framework may be eventually proven futile. Diffusion is too slow to ensure observed characteristic times of the establishment of morphogenetic gradients \cite{rev16}, and morphogens may be delivered in a more sophisticated way through cell extensions specialised in transporting them to cells \cite{Wolpert07}. This implies more intricate and specialised interactions that determine  positional information.  

Mechanical regulation may complement chemistry in the morphogenesis as well as in the functioning of living cells and tissues \cite{Belintsev, Taber}. The perspective of applying general equations of continuous mechanics in their more sophisticated forms, as in the active gel \cite{gel} and poroelastic \cite{poro} theories, is certainly appealing to the physicists. Chemo-mechanical interactions involve, however, complex specific mechanisms, and the applicability of continuum theories is impeded by the crowded and irregular microstructure of living cells.

\section{The devil in details}

The generalising attitude of 20th century physicists has given way in the new century to focusing on details. Tens of thousands researchers, sustained by generous funding, apply their ingenuity and modern technology to study detailed genetic expression details of the famous \emph{Drosophila}, with the attendance of ``Fly" meetings perhaps exceeding those of the American Physical Society. Some features of the morphogenetic process are amazingly conserved, from \emph{Drosophila} to higher animals, but curiosity and inertia may also drive researchers into thorough studies with irrelevant results. Details are essential when studying human development and physiology, even at the price of sacrificing our mammal relatives to save human lives, but not every detailed study of the proteins involved in the formation mechanism of a particular feature of the fly anatomy would contribute to either practical or existential knowledge. 

Of course, detailed studies help to elucidate general principles as well; in particular, only in this way the relative role of diffusion and active transport in shaping morphogenetic patterns can be understood. More insight is promised by \emph{in vitro} studies in controlled artificially engineered environment \cite{vitro}. \emph{In silico} studies are also actively pursued but their results should be viewed with caution, inasmuch as mathematical models are apt to produce results superficially similar to observations even when their foundations are far from reality. The detailed view tends to be infinitely complex. In the end, by Wolpert's \cite{Wolpert16} admission, ``we still do not know the molecular basis of positional information [...], nor do we have convincing evidence of how positional values are specified or interpreted. Even the role of diffusion in morphogens is unclear". The other cited review \cite{Schier} ends at a similar note.

It is necessary to know what do we really need to know. In Schr\"odinger's words \cite{Schr}, we really know only what we can make. Even this may cease to be true in the age of artificial intelligence (AI) but details certainly matter as manufacturing adopts biomimetic principles of pattern formation governed by signalling, replacing or complementing the assembly methods (which also encompass, alongside traditional manufacturing, such modern processes as lithography or 3D printing). 

A promising direction is the use of stimuli-responsible soft materials, such as hydrogels and nematic elastomers, which are distinguished by feedback interactions between chemical and environmental signals and mechanical properties --  similar to natural materials but on a far lower level of complexity. 
The BZ reaction in a hydrogel causes its mechanical oscillations \cite{Balazs,Borckmans}. The orientation texture of nematic elastomers can be prepared in such a way that they acquire a desired shape upon transition to the isotropic state \cite{Broer,White,face}. Reshaping can be also induced in biomimetic fashion by externally imposed dopant gradients \cite{SM17}. Repeated reshaping has been use to simulate \cite{ZP16} and reproduce in the laboratory \cite{Hao} artificial crawlers.

One can expect that intelligent design, either by humans or AI, will come to simpler and more rational  (if not superior) solutions than blind Darwinian search, and is likely to rely on electric rather than chemical signalling. We do not expect that drones will ever be manufactured as insects are, any more than airplanes as birds or computers as brains, notwithstanding the wonders of natural design -- but when sophistication of intelligently designed autonomous systems reaches the level of their living counterparts, their production process and behaviour may also become too complex to be reduced to a few elegant general principles.  

\emph{Acknowledgement} This minireview has been submitted by invitation by Prof. M.~Tlidi to the issue of  Proceeding of Royal Society A in memory of Ilya Prigogine, and rejected by anonymous referees for their personal reasons.

\end{document}